%% file: main.tex
\documentclass[lettersize,journal]{IEEEtran}
\usepackage{cite}
\usepackage{amsmath,amsfonts}
\usepackage{booktabs} 
\usepackage[caption=false,font=normalsize,labelfont=sf,textfont=sf]{subfig}
\usepackage{textcomp}
\usepackage{stfloats}
\usepackage{url}
\usepackage{verbatim}
\usepackage{graphicx}
\usepackage{algorithm} 
\usepackage{algpseudocode} 
\usepackage{xspace}
\usepackage{colortbl}
\usepackage{multirow}
\usepackage[inkscapelatex=false]{svg}
\usepackage{tcolorbox}
\usepackage{xcolor}
\usepackage{xfp}
\usepackage{microtype}
\usepackage{enumitem}
\usepackage{subfiles}
\usepackage{listings}
\usepackage{url}
\usepackage{pifont}
\usepackage{cleveref}
\newcommand{\ourtool}{DeepFWI\xspace}
\newcommand{\etal}{\textit{et al.}\xspace}

\newcommand{\ourdataset}{BSWarnings\xspace}
\usepackage{arydshln}
\newcommand{\revise}[1]{\textcolor{black}{#1}}
\newcommand{\revised}[1]{\textcolor{black}{#1}}

\definecolor{codegreen}{rgb}{0,0.6,0}
\definecolor{codegray}{rgb}{0.5,0.5,0.5}
\definecolor{codepurple}{rgb}{0.58,0,0.82}
\definecolor{shallowred}{rgb}{1,0.8,0.8}

\definecolor{codegreen}{rgb}{0,0.6,0}
\lstdefinelanguage{diff}{
    frame=single,
    basicstyle=\ttfamily\scriptsize\bfseries, 
    columns=fullflexible, 
    morecomment=[f][\color{red}]{---}, 
    morecomment=[f][\color{codegreen}]{+++},
    morecomment=[f][\color{red}]{-\ },
    morecomment=[f][\color{codegreen}]{+\ },
    morecomment=[f][\color{blue}]{@@},
    breakatwhitespace=false,
    breaklines=true, 
    postbreak=\mbox{\textcolor{gray}{$\hookrightarrow$}\space},
    captionpos=b,
    rulecolor=\color{black},
    keepspaces=true,
    numbers=left,
    numbersep=5pt,
    showspaces=false,
    showstringspaces=false,
    framexleftmargin=2mm, 
    framexrightmargin=2mm,
    framextopmargin=1mm,
    framexbottommargin=1mm,
    showtabs=false,
    tabsize=2,
    escapeinside={(*@}{@*)},
    framexleftmargin=12pt,
    xleftmargin=10pt,
}
\definecolor{codegreen}{rgb}{0,0.6,0}
\definecolor{codegray}{rgb}{0.5,0.5,0.5}
\definecolor{codepurple}{rgb}{0.58,0,0.82}
\definecolor{shallowred}{rgb}{1,0.8,0.8}

\lstdefinestyle{mystyle}{
    language=Java,
    commentstyle=\color{codegreen},
    keywordstyle=\color{blue},
    stringstyle=\color{codepurple},
    basicstyle=\ttfamily\scriptsize\bfseries,
    breakatwhitespace=false,
    breaklines=true,
    postbreak=\mbox{\textcolor{gray}{$\hookrightarrow$}\space},
    captionpos=b,                    
    keepspaces=true,                 
    numbers=left,                    
    numbersep=5pt,                  
    showspaces=false,                
    showstringspaces=false,
    frame=single,                    
    framexleftmargin=0mm,            
    framexrightmargin=0mm,           
    framextopmargin=0mm,             
    framexbottommargin=0mm,    
    showtabs=false,                  
    tabsize=2,
    escapeinside={(*@}{@*)},
    framexleftmargin=12pt,
    xleftmargin=10pt,
}

\ifCLASSINFOpdf
\else
\fi

\begin{document}
%
\title{\revised{DeepFWI: Identifying Bug-Sensitive Warnings with Multi-Modal Code-Warning Semantics}}
%
%
%

\author{Han~Liu,
        Jian~Zhang,
        Cen~Zhang,
        Xiaohan~Zhang,
        Kaixuan~Li,
        Sen~Chen,
        Shang-Wei~Lin,
        Yixiang~Chen,
        Xinghua~Li,
        and~Yang~Liu,~\IEEEmembership{Senior Member,~IEEE}

\thanks{
{
\IEEEcompsocthanksitem Han Liu, Sen Chen is with College of Cryptology and Cyber Science, Nankai University, Tianjin 300350, China. (E-mail: \{hanliu, senchen\}@nankai.edu.cn)
\IEEEcompsocthanksitem Jian Zhang is the corresponding author. He is with School of Software, Beihang University, Beijing 100191, China. (E-mail: zhangj\_cs@buaa.edu.cn)
\IEEEcompsocthanksitem Cen Zhang, Kaixuan Li, and Yang Liu are with Nanyang Technological University, Singapore 639798.
(E-mail: \{cen001, kaixuan.li, yangliu\}@ntu.edu.sg)
\IEEEcompsocthanksitem Xiaohan Zhang is with the School of Computer Science, Shanghai Jiaotong University, Shanghai 200062, China. (E-mail:  xhzhang1@sjtu.edu.cn)
\IEEEcompsocthanksitem Shang-Wei Lin is with Singapore Institute of Technology, Singapore 828608. (E-mail: shangwei.lin@singaporetech.edu.sg)
\IEEEcompsocthanksitem Yixiang Chen is with the Shanghai Key Laboratory of Trustworthy Computing, East China Normal University, Shanghai 200062, China. (E-mail: yxchen@sei.ecnu.edu.cn)
\IEEEcompsocthanksitem Xinhua Li is with Xidian University, Xi'an 710126, China. (E-mail:  xhli1@mail.xidian.edu.cn)
} 
}
}

%
%

\markboth{Journal of \LaTeX\ Class Files,~Vol.~14, No.~8, August~2021}%
{Han Liu \MakeLowercase{\textit{et al.}}: DeepFWI: Identifying Bug-Sensitive Warnings with Multi-Modal Code-Warning Semantics}

%



\maketitle



\input{sections/0-abstract}

\maketitle
\begin{IEEEkeywords}
Deep learning, warnings, automated static analysis tool, bug-sensitive.
\end{IEEEkeywords}

%

\input{sections/1-Introduction}
\input{sections/2-preliminary}
\input{sections/3-methodology}
\input{sections/4-evaluation}
\input{sections/5-threats}
\input{sections/6-relatedwork}

\input{sections/7-conclusion}

\section*{ACKNOWLEDGMENT}
We thank the Associate Editor and the anonymous reviewers for their insightful comments and constructive suggestions.

\ifCLASSOPTIONcaptionsoff
  \newpage
\fi



\bibliographystyle{IEEEtran}
\bibliography{main}

\end{document}

%% file: sections/0-abstract.tex
\begin{abstract}
Static analysis tools have evolved over time to assist in detecting bugs. However, the excessive false warnings can impede developers' productivity and confidence in the tools. Previous research efforts have explored learning-based \revise{approaches} to identify bug warnings. Nevertheless, their coarse granularity, focusing on either long-term warnings or function-level alerts, are insensitive to individual bugs. Also, they rely on manually crafted features or solely on source code semantics, which is inadequate for effective learning.
In this paper, we propose \ourtool, a learning-based approach that identifies bug-sensitive warnings at a fine-grained granularity. Specifically, we design a novel LSTM-based model that captures multi-modal semantics of source code and warnings from automated static analysis tools (ASATs) and highlights their correlations with cross-attention. To tackle the data scarcity of training and evaluation, we collected a large-scale dataset of 280,273 warnings. We conducted extensive experiments on the dataset to evaluate \ourtool. The experimental results demonstrate the effectiveness of our approach, with an F1-score 67.06\% for confirming true warnings in a finer-grained manner, significantly outperforming all baselines. Additionally, to validate the practicality of DeepFWI from the perspective of developers, we applied DeepFWI to four popular open-source projects. Our \revise{approach} filtered out the vast majority of warnings, while still successfully surfacing 25 true bug-related warnings that were confirmed through manual analysis.
\end{abstract}

%% file: sections/1-Introduction.tex
\section{Introduction}\label{intro}
\IEEEPARstart{A}{utomated} static analysis tools (ASATs) are widely adopted in industry for their ability to detect bugs at low cost and without code execution~\cite{yedida2023find}. However, a longstanding challenge remains: ASATs often produce an excessive number of warnings, the majority of which are ultimately irrelevant or false positives. Recent studies show that up to 90\% of ASAT warnings may be false positives~\cite{kang,issta23}, overwhelming developers and hindering effective bug detection.

To mitigate this flood of warnings, researchers have developed learning-based approaches to prioritize or filter ASAT outputs. Two main directions have emerged: (1) \textit{Ranking} warnings to highlight those most likely to be fixed, and (2) \textit{Identifying} bug-related warnings for focused resolution. However, existing ranking approaches~\cite{rank1,rank2,rank3,rank4} typically operate on groups of warnings and leave open questions about individual warning prioritization. Developers are often left unsure of which specific warnings demand immediate attention. Therefore, we focus this work on the finer-grained problem of identifying which individual warnings are truly related to bugs.

A major thread in prior work is the notion of \textit{actionable warnings}: those ASAT warnings that are likely to be resolved during the software lifecycle~\cite{Ruthruff,Yang,Heckman,HECKMAN2011363,golden,Hanam,Williams,Yüksel}. Classical approaches use machine learning classifiers built on manually engineered features of warnings, code, history, and file attributes~\cite{Hanam,Williams,golden,Yüksel}. For example, Wang~\etal identified 23 ``Golden Features'' that improved actionable warning detection~\cite{golden}. Yet, recent studies highlight that such \revise{approaches} are vulnerable to data leakage and duplication, undermining their reliability~\cite{kang,yedida2023find}.

An alternative direction leverages deep learning for bug prediction at the function level~\cite{Kharkar}, often using pre-trained Transformer models. While these \revise{approaches} can boost precision, they may ignore the specific semantics of warning messages and treat all warnings in a function uniformly, making it hard to distinguish between true and false positives on a per-warning basis. More recently, Large Language Models (LLMs) like GPTs have been used via prompt engineering~\cite{Wen}, but these \revise{approaches} are expensive to scale and susceptible to prompt sensitivity and hallucinations.

Despite progress, several key limitations persist in the state of the art:
\begin{description}
    \item[\textbf{\revise{Limitation} 1}] \textbf{Ambiguous Granularity.} For actionable warnings detection, existing studies define them over coarse time frames (e.g., months between commits), making it unclear whether a warning's disappearance truly signals a bug fix or is due to unrelated code changes such as refactoring~\cite{golden,Yang,kang,yedida2023find}. As for function-level bug prediction~\cite{Kharkar}, it also fails to distinguish between multiple warnings within the same function.
    \item[\textbf{\revise{Limitation} 2}] \textbf{Limited Semantics.} \revise{Feature-based approaches (i.e.,~\cite{golden,Yang,kang,yedida2023find,FRUGAL})} often miss the rich semantics embedded in both code and warning messages, limiting their ability to distinguish true from false positives. Function-level models~\cite{Kharkar} typically aggregate the entire function into a single representation, which can dilute the specific signals associated with a warning. Additionally, they ignore the rich semantic cues present in warning messages and other relevant code definitions (e.g., class fields), leading to suboptimal performance in identifying bug-sensitive warnings.
    \revise{The LLM-based approach~\cite{Wen} is} computationally expensive when applied to large-scale data and prone to irrelevant outputs due to prompt variability and hallucinations.
    \item[\textbf{\revise{Limitation} 3}] \textbf{Lack of Scale and Generalizability.} Most prior work~\cite{golden,Yang,kang} is evaluated on limited datasets, which are often small in scale and lack diversity. This leaves open questions about scalability and generalizability to diverse, real-world projects.
\end{description}

To address these challenges, we introduce the notion of \textit{bug-sensitive warnings}: \revise{individual ASAT warnings that appear in a buggy version of code and disappear as a direct result of the corresponding bug-fix commit} (formally defined in~\Cref{problem_definition}). Unlike generic actionable warnings, which may reflect code changes over long periods and uncertain scope, bug-sensitive warnings are tightly coupled to concrete bug fixes, providing a more precise and timely signal for developers. By focusing on the disappearance of warnings directly tied to bug-fixing commits, we minimize ambiguity caused by unrelated code modifications and improve the granularity of bug detection.

However, automatically identifying bug-sensitive warnings still presents two core challenges:
\begin{description}
    \item[\textbf{Challenge 1}] \textbf{Dataset Scarcity.} There is no existing large-scale dataset of bug-sensitive warnings suitable for training and evaluation. To bridge this gap, we construct \textbf{BSWarnings}, the largest warning-related dataset to date, by mining 607,259 bug-fixing commits across 32,100 open-source projects. Using SpotBugs~\cite{Spotbugs}, we systematically pair warnings before and after bug-fix commits, applying rigorous filtering to eliminate spurious cases and deduplication. \textbf{BSWarnings} contains 280,273 total warnings, including 20,100 bug-sensitive warnings, validated via random sampling at over 92\% accuracy, with a confidence level of 95\%. Notably, BSWarnings exceeds existing benchmarks by an order of magnitude (10 times)~\cite{golden}. To achieve this scale, we executed over 1.2 million compilation and analysis tasks, totaling more than 5,000 hours of distributed computation on three servers.
    \item[\textbf{Challenge 2}] \textbf{Multi-Modal Semantics.} Bug-sensitive warning identification requires capturing multi modal semantics: the joint representation of several distinct but complementary information sources associated with each warning, for example, the source code context where the warning arises, the natural language warning message, and structured warning attributes (e.g., warning type, rank). To address this, we propose \textbf{DeepFWI}, a novel neural architecture that advances the state of static analysis warning identification in several key ways: 
    \begin{itemize}
    \item \textbf{Multi-modal, fine-grained representation:}  
    DeepFWI processes both source code and warning messages as first-class, complementary modalities. Unlike previous models that treat warnings in aggregate or ignore warning text, DeepFWI encodes each warning individually, integrating not only the relevant code context but also the structured attributes and rich natural language of warning messages.
    \item \textbf{Cross-attention mechanism for code-warning correlation:}  
    We introduce a cross-attention module that explicitly models the semantic alignment between code slices and warning messages. This allows DeepFWI to focus on the most relevant code regions for each warning, capturing subtle interactions that prior sequence models or simple pooling approaches cannot.
    \item \textbf{Context precision via warning-aware slicing:}  
    DeepFWI incorporates a slicing strategy, extracting code slices tightly linked to each warning. This reduces noise, improves context quality, and ensures the model bases predictions on the most relevant program elements.
    \end{itemize}
\end{description}

In summary, DeepFWI represents a significant advancement \revise{in bug-sensitive warnings identification generated by ASATs} by combining a large-scale, rigorously validated dataset of bug-sensitive warnings with a novel neural architecture that captures multi-modal semantics at fine granularity. Table~\ref{tab:comparison} presents a structured comparison of DeepFWI versus state-of-the-art (SOTA) techniques across four critical dimensions: granularity, input modality, and computational cost. As shown, while feature-based approaches~\cite{golden,Yang,kang,yedida2023find,FRUGAL} rely on coarse-grained manual features and function-level models~\cite{Kharkar} lack specific warning context, DeepFWI uniquely targets fine-grained code slices. Furthermore, unlike \revise{the LLM-based approach}~\cite{Wen} which \revise{incurs} high inference costs and \revise{risks} hallucinations, DeepFWI leverages a specialized cross-attention mechanism to align code and warning semantics efficiently. This comparison highlights the specific research gap DeepFWI aims to fill: the need for a low-cost, fine-grained, and semantically aware solution.

\begin{table*}[t]
\caption{Comparison of DeepFWI with State-of-the-Art Approaches}
\label{tab:comparison}
\centering
\scalebox{1.05}{
\begin{tabular}{l|ccc}
\toprule
\textbf{Approach} & \textbf{Granularity} & \textbf{Input Modality} & \textbf{Cost} \\ \midrule
Feature-based~\cite{golden,Yang,kang,yedida2023find,FRUGAL} & Coarse Time Frames with Warnings & Manual Features & Low \\
Function-level~\cite{Kharkar} & Function &  Implicit & Medium \\
LLM-based~\cite{Wen} & Function & Code + Prompt  & High \\ \midrule
\textbf{DeepFWI (Ours)} & \textbf{Bug-Sensitive Warnings} & \textbf{Multi-modal (Code+Messages+Sliced IR+Attributes)}  & \textbf{Low} \\ \bottomrule
\end{tabular}}
\end{table*}

We comprehensively evaluate DeepFWI against ten baselines, demonstrating its superior precision (at least +9.0\%) and F1-score (at least +13.2\%) compared to SOTA \revise{approaches}, including feature-based~\cite{golden,Yang,kang,yedida2023find,FRUGAL}, function-level~\cite{Kharkar}, LLM-based \revise{approach}~\cite{Wen}, and several straightforward \revise{solutions}~\cite{graphcodebert,vaswani2017attention,biLSTM,brown1992class}. 
\revise{Specifically, in terms of F1-score, \ourtool outperforms feature-based approaches~\cite{golden,Yang,kang,yedida2023find,FRUGAL} by 15.81\%, the function-level approach~\cite{Kharkar} by 14.01\%, and the LLM-based approach~\cite{Wen} by 53.78\%. When evaluated against straightforward baselines, \ourtool continues to show superiority, surpassing N-gram~\cite{brown1992class}, Transformer~\cite{vaswani2017attention}, GraphCodeBERT~\cite{graphcodebert}, and LSTM~\cite{biLSTM} models by 17.14\%, 15.55\%, 16.42\%, and 13.17\%, respectively. Overall, DeepFWI establishes a clear performance advantage, with F1-score margins ranging from 13.2\% to 57.6\%.}

Unlike standard Transformers which impose strict length limits (e.g., 512 tokens) to maintain efficiency, and modern LLMs which, despite handling long contexts, suffer from substantial computational overhead and slow inference speeds, for instance, the inference time of the modern Qwen2.5-0.5B model is approximately 180 times that of \ourtool provide an optimal trade-off. They efficiently encode multi-dimensional information and flexibly accommodate longer inputs, preserving critical context while avoiding the heavy memory usage and energy consumption inherent to large-scale Transformer architectures. Our ablation studies (Section IV-D2) further demonstrate that LSTM-based architectures, especially when enhanced with cross-attention, are effective at capturing the sequential and localized dependencies between code slices and warning messages in this task. We also discuss these trade-offs in detail and provide empirical comparisons with state-of-the-art Transformer and LLM baselines.

To assess real-world applicability, we deploy DeepFWI on four popular open-source Java projects (\textit{itext7}, \textit{metadata-extractor}, \textit{poi}, \textit{pdfbox}). While DeepFWI filters out over 92\% of potential false alarms—greatly reducing developer workload—it still successfully identifies 25 previously unknown bugs, 4 of which were confirmed and fixed by project maintainers. We also analyze the risk of filtering out true positives and discuss its practical implications.

In summary, this paper makes the following contributions:
\begin{itemize}
    \item We present \textbf{BSWarnings}, a rigorously validated dataset of 280,273 warnings, including 20,100 bug-sensitive warnings, mined from over 600K bug-fixing commits.
    \item We propose \textbf{DeepFWI}, a novel neural model for fine-grained, warning-level bug identification, integrating multi-modal semantics from source code and warning messages using cross-attention.
    \item We provide an extensive evaluation that DeepFWI outperforms ten SOTA baselines and generalizes to new projects, supporting practical adoption.
    \item All data and code are publicly released to future research~\cite{website}.
\end{itemize}

%% file: sections/2-preliminary.tex
\section{Preliminary}
\subsection{Automated Static Analysis Tools}\label{describe}
Automated Static Analysis Tools (ASATs), popular in software development, are designed to analyze code for bugs and code smells without execution. Notable examples include SonarQube~\cite{Sonarqube}, SpotBugs~\cite{Spotbugs} (formerly FindBugs~\cite{findbugs}), and Infer~\cite{infer}. 
In particular, SpotBugs~\cite{Spotbugs} is an open-source static code analysis tool utilized in software development for bug detection in Java code. As the direct successor to FindBugs~\cite{findbugs}, it is among the most highly regarded tools in its domain. SpotBugs carries forward the same fundamental technology but with continuous support and improvements.

\begin{figure}
\begin{lstlisting}[language=diff]
private static StringBuilder sendRequest(...) throws Exception{
      ...
  if (HttpMethods.POST.equals(method)) {
       ...
-   connection.setRequestProperty(REQUEST_HEADER_CONTENT_LENGTH, Integer.toString(data.getBytes().length));
+ //connection.setRequestProperty(REQUEST_HEADER_CONTENT_LENGTH, Integer.toString(data.getBytes().length));
       ...
    }
  ...
  InputStream is = connection.getInputStream();
  BufferedReader rd = new BufferedReader(new InputStreamReader(is));
     ...
  }
\end{lstlisting}
\caption{Simplified code snippets from DongTai-agent-java (Commit 2822534) that motivate our work}
\label{code1}
\end{figure}

While ASATs are effective in bug detection during software development stages~\cite{6606642,9099449,9124719,9736606}, they depend on human-made rules or established patterns to identify known errors. The tendency of ASATs to generate an overwhelming number of warnings to cover potential bugs limits their usefulness in large software projects~\cite{Johnson,Bielik,issta23,Ayewah}. To mitigate the occurrence of false alarms in ASATs, various methodologies have been proposed to identify truly beneficial warnings. One focal point of these studies is actionable warnings, i.e., warnings that developers would rectify over time. Such studies contemplate multiple features of the warnings and the accompanying code~\cite{Hanam,Williams,golden,Yüksel}.
However, there exist issues like data leakage and duplication that severely undermine the reliable identification of truly actionable warnings~\cite{kang}. 
Moreover, actionable warnings may not reflect the remediation of bugs since it is based on the long span of commit history where the source code evolves greatly.
Alternatively, Kharkar \etal utilized a learning \revise{approach} to minimize false positives~\cite{Kharkar}. They apply CodeBERTa as the foundation model to identify the function of related warnings whether there is a bug and show an improved accuracy. Clearly, this \revise{approach} operates at a coarse granularity that considers only the entire function.

\subsection{Actionable Warnings}\label{actionable_warnings}
In the context of static analysis, an actionable warning refers to a warning reported by an ASAT that is deemed relevant and meaningful for developers to address during software maintenance~\cite{golden,Yang,kang,yedida2023find}. Unlike generic warnings that may include minor code style suggestions or low-impact issues, actionable warnings highlight potential defects or code problems that are likely to be fixed by developers over the course of normal development. 

Formally, a warning is considered actionable if it corresponds to an issue that is subsequently resolved—typically within a defined time frame (e.g., 3 months)—through code changes such as bug fixes or targeted improvements. Actionable warnings are thus distinguished by their practical impact: they prompt corrective action and have a higher likelihood of being addressed compared to non-actionable or spurious warnings. Prior work~\cite{golden,Yang,kang} has shown that identifying actionable warnings can significantly reduce developer workload by filtering out noise and focusing attention on the most critical alerts.

\subsection{Motivating Example}\label{motivate}

When dealing with vast and numerous projects, developers often find themselves overwhelmingly burdened by a large number of warnings reported by ASATs~\cite{issta23}. Hence, it is necessary to reduce the false alarms of the warnings. However, existing work can easily encounter problems due to the predefined features or the coarse granularity. They cannot effectively distinguish the individual warnings, which may result in mislabeling.

The following commit\footnote{\url{https://github.com/HXSecurity/DongTai-agent-java/commit/2822534}} is from the project DongTai-agent-java of organization HXSecurity. As shown in~\Cref{code1}, it fixed a bug caused by \textit{data.getBytes()} when the platform's default charset is not compatible with the data being sent.
Since the \texttt{CONTENT\_LENGTH} header is not always necessary in a POST request, the fixed version has commented Line 5 out, and let the request automatically handle the content length.

SpotBugs reported a warning related to reliance on default encoding at line 5, which disappeared in the revised version. As such, it qualifies as a bug-sensitive warning. Meanwhile, SpotBugs issued a similar warning at line 10, involving the same function. Here, however, the \textit{InputStream (is)} involves raw byte data obtained from the server response, which is then decoded into characters using the default charset. This operation actually functions correctly because the server's response is encoded using a charset compatible with the default. Accordingly, we identified this as a bug-insensitive warning. Given that the two warnings originated from the same function, existing function-level learning-based warning verification \revise{approaches} and feature-based \revise{approaches} would consider the same input, potentially leading to mislabeling. Hence, to distinguish these bug-sensitive warnings from the bug-insensitive warnings and lower human effort when finding bugs using ASATs, a finer granularity \revise{approach} for the warning identification is imperative.

\subsection{Problem Definition}\label{problem_definition}

Our task is to facilitate developers to identify the bug-sensitive warnings, allowing us to reduce the burden on developers of sifting through potentially large numbers of irrelevant warnings, and instead focus their efforts on addressing bugs that have been identified with high confidence. We formally define the task as follows.

In the context of static analysis, where the process maps a codebase $\mathcal{C}$ to a set of warnings $\mathcal{W}$ (i.e., $\text{SA}: \mathcal{C} \rightarrow \mathcal{W}$), a bug-sensitive warning is defined to rigorously capture warnings causally linked to actual defects. Formally, given a historical dataset $\mathcal{H}$ of validated bug-fix pairs $(C_b,C_f)$, where $C_b$ represents buggy code and $C_f$ represents fixed code, we define a set of bug-sensitive warnings $W_b \subseteq \mathcal{W}$ as follows:

\noindent A warning $w \in \mathcal{W}$ is classified as \textit{bug-sensitive} if:
\begin{equation}
    w \in W_b \iff \exists (C_b, C_f) \in \mathcal{H} :
    \begin{cases}
    w \in SA(C_b) \wedge w \notin SA(C_f), \\
    L(w, C_b) \cap \Delta(C_b, C_f) \neq \emptyset.
    \end{cases}
\end{equation}
\noindent
where:
\begin{itemize}
\item $L(w, C_b)$ maps $w$ to code locations (i.e., lines) in $\mathcal{C}$ where $w$ is triggered,
\item $\Delta(C_b, C_f)$ denotes code modifications between buggy code$C_b$ and fixed code$C_f$.
\end{itemize}

Given a codebase $\mathcal{C} = \{c_1, c_2 ,..., c_N\}$ and its associated static analysis warnings $\mathcal{W} = \{w_1,w_2,...,w_N\}$, the task is to classify each warning $w_j \in \mathcal{W}$ as either bug-sensitive or bug-insensitive. Formally, this is framed as a supervised learning problem where a classifier $\text{F}$ learns the discriminative mapping:
\begin{equation}
    \mathbf{F} : \mathcal{C} \times \mathcal{W} \to \{0,1\}, \quad \text{such that } \mathbf{F}(c_i, w_j) = y_{ij}
\end{equation}
Here, $y_{ij} \in \{0,1\}$ is a binary label where $ y_{ij} = 1 $ indicates that warning  $w_j$  in code snippet $c_i$ is bug-sensitive, and  $y_{ij} = 0$ otherwise.

\revise{Given the variations across programming paradigms and languages}, our approach focuses on Java and SpotBugs following the existing research~\cite{golden,Kharkar,Yang}. To clarify the terminology, we use ``function'' to describe our methodology to present a comparable discussion with existing work~\cite{golden,Kharkar,Yang} using ``function'', while ``method'' is used in some Java-specific contexts to refer to Java methods. 

%% file: sections/3-methodology.tex
\section{Methodology}
\subsection{Overview}
In this section, we introduce \ourtool, a deep learning-based (DL-based) approach to identify warnings of bugs for ASATs.
Different from existing feature-based~\cite{golden,Yang} and learning-based approaches~\cite{Kharkar}, \ourtool is designed to automatically capture more comprehensive semantic information from the code, warnings, and their interaction, applying it across a broader scope and with finer granularity warning identification. 

\subsection{Data Collection}
\begin{figure}
\centerline{{\includegraphics[width=0.48\textwidth]{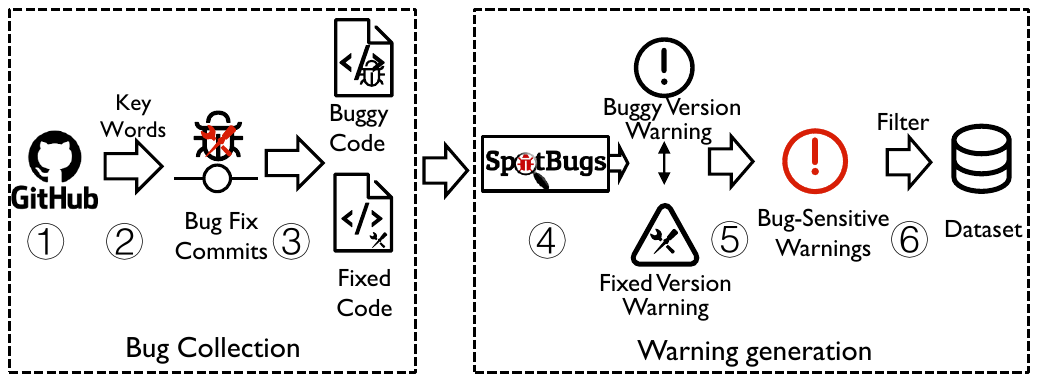}}}
\caption{The pipeline for collecting bug-fix commits and generating bug-sensitive warning datasets.}
\label{datacollectionprocess}
\end{figure}
\label{datacollection}
Considering that current datasets of warnings, such as ~\cite{golden}, are limited by coarse granularity, either in the dimension of time (i.e., actionable warnings) or code (i.e., function level), they do not align with our requirement for bug-sensitive warnings identification, which demands a finer granularity. In addition, the scale of the existing datasets is limited to the specific projects. Hence, 
to construct an automated approach that identifies warnings of bugs via deep learning, our initial step involves creating a dataset, \ourdataset, gathering a large number of bug-sensitive warnings. The overall data collection process is illustrated in~\Cref{datacollectionprocess}, which consists of six main steps:

\begin{enumerate}
\item[\textbf{\ding{172}}] \textbf{Project Selection:} We begin by searching for as many Java projects as possible. \revise{To ensure dataset quality, we filtered projects by selecting those in the top 0.3\% (about 32,100 Java projects) based on GitHub stars to ensure we focused on popular and actively maintained repositories. The 0.3\% threshold was chosen to balance the need for a manageable dataset size while still capturing a diverse range of high-quality projects. Below this threshold, we observed a significant drop in project activity and maintenance, which could affect the reliability of our analysis.}

\item[\textbf{\ding{173}}] \textbf{Bug-Fixing Commit Identification:} We identify bug-fixing commits within these projects by applying keyword filtering to commit messages, following prior work~\cite{keyword}. Specifically, to be identified as such, a commit message must contain the word ``fix'' along with other bug-related words such as ``bug'', ``defect'', ``error'', ``fault'', ``issue'', or bug classification words like ``null pointer dereference'', ``resource leak'' etc. Using this filtering process, we collected a total of 607,259 related commits. 

\item[\textbf{\ding{174}}] \textbf{Buggy and Fixed Code Extraction:}  For each identified commit, we clone the project and obtain both the buggy (pre-fix) and fixed (post-fix) versions of the code.

\item[\textbf{\ding{175}}] \textbf{Warning Generation:} We compile both code versions and run SpotBugs to generate static analysis warnings for each.

\item[\textbf{\ding{176}}] \textbf{Bug-Sensitive Warning Identification:}We compare the warnings generated from the relevant files in both the buggy and fixed versions of the commits. If the warning disappears in the fixed version, we mark it as a bug-sensitive warning. For deleted functions, if the corresponding functions are missing in the fixed version, then by our definition, $\Delta(C_b, C_f)$ is empty. Consequently, we do not consider the warning to be bug-sensitive—instead, we designate it as bug-insensitive because it is not correlated with the specific bug. Similarly, newly added functions are excluded from consideration. Since there is no corresponding buggy version available for comparison, the warnings generated by these newly added functions fail to meet the second condition, which requires that $L(w, C_b) \cap \Delta(C_b,C_f)$ be non-empty.

\item[\textbf{\ding{177}}] \textbf{Deduplication and Final Dataset Construction:} Finally, we remove duplicate and bug-insensitive warnings, resulting in \ourdataset, which contains 280,273 warnings, including 20,100 bug-sensitive warnings.
\end{enumerate}

Notably, \ourdataset represents the largest collection of warning-related data to date, exceeding previous benchmarks by an order of magnitude (10 times)~\cite{golden}. To achieve this scale, we executed over 1.2 million compilation and analysis tasks, totaling more than 5,000 hours of continuous distributed computation across three high-performance CPU servers, \revise{each equipped with dual Intel x64 48-core CPUs and 256 GB of RAM with Ubuntu 18.04 Operating System}.

\subsection{Fine-grained Warning Identification}
\begin{figure*}
\centering
\includegraphics[width=0.98\textwidth]{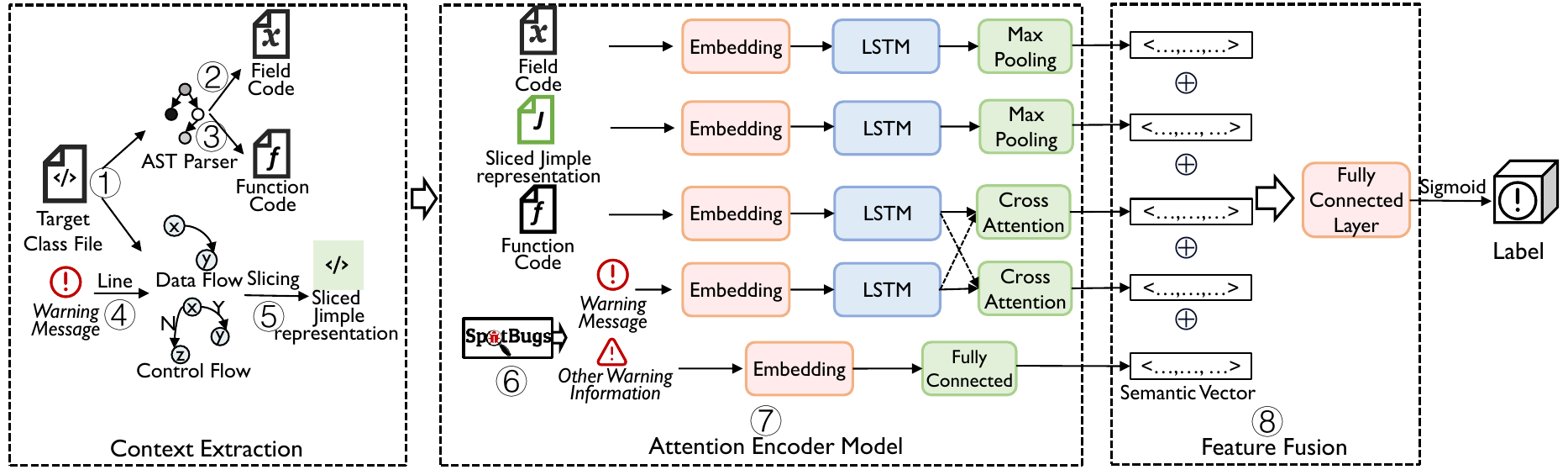}
\caption{The architecture of DeepFWI: attention-based warning identification model.}
\label{framework}
\end{figure*}

Now we present our DL-based approach to bug-sensitive warning identification. 
Our approach includes four main components, i.e., code context extraction, warning information extraction, attention encoder model, and feature fusion. The framework of our approach is shown in ~\Cref{framework}. Firstly, we extract the code context and warning information from the target class reported by specific warnings. Then, based on the code context and warning information, we design an attention encoder model on them to obtain a unified vector representation. The vector is finally used to determine whether the warning is bug-sensitive or not. 
\subsubsection{Code Context Extraction \textnormal{\textbf{(\ding{172}--\ding{176}} in~\Cref{framework})}}
The step of code context extraction consists of three key parts: warning function extraction ((\ding{172} and \ding{174} in~\Cref{framework})), field of class extraction ((\ding{172} and \ding{173} in~\Cref{framework})), and warning-aware slicing ((\ding{175} and \ding{176} in~\Cref{framework})).
The warnings provide the potential bugs present in a specific class and function with the specific lines. Due to the multi-modal sources involved in one warning, such as code, messages, and rules, it becomes necessary to extract sufficient and precise context for the model to effectively identify the warnings. Considering that the length of input for a neural model is typically limited, our context extraction of the code is within the target class. 

\noindent\textbf{\textit{Warning Function Extraction} \textbf{(\ding{172} and \ding{174} in~\Cref{framework})}}:
To provide more related information for the network, we first focus on the function (i.e., method) on which the specific warning is reported. As described before, each warning points to the class where the bugs may exist, so we locate the file for further extraction. However, due to the likelihood that the warning is missing the function name, in addition, there may be function overloading in the specific class, we extract the function with the following criteria: (1) If the warning is reported with lines, we use the lines to locate the unique function. If the lines are not a specific function, we recognize the context of this warning is the whole class. (2) If the warning is reported with a function name, we locate all the functions with the same name as the context. Otherwise, we use the class as the context of the warning. In this step, we use ANTLR4~\cite{antlr4} to build the abstract syntax tree (AST) for the buggy file and then obtain the function. If we cannot find the function corresponding to the line number, or if there is no line number available, we use the entire class content to fill in this input.

\noindent\textbf{\textit{Field of Class Extraction} \textbf{(\ding{172} and \ding{173} in~\Cref{framework})}}: 
In addition, considering the characteristics of the Java language, Java fields, i.e., member variables in a class~\cite{javadoc}, which are declared inside a class, but outside any method, constructor, or block, can be related to all functions. 
They are object-specific and have potentially different values for different objects of the same class. Fields can be categorized as public, private, or protected, which are visibility modifiers that define how and from where these fields can be accessed. This external context might also hold key clues related to various potential errors. For instance, issues like incorrect field initialization could lead to unexpected application behavior, while null pointer exceptions might arise from fields not being properly instantiated. Therefore, we extract the field code in the class to provide more information for the model. In this work, field code refers to the segment of source code that is dedicated to declaring class-level fields, including field names, modifiers, types, and initial values or assignments. For example, in a declaration like \texttt{private int counter = 0;}, the field code would include the modifier (private), the type (int), the field name (counter), and the initializer (0). To extract the field code, we start with the AST constructed by the code file. Then, we locate the root of the class and traverse the class. If the node is the definition of variables and does not belong to any function method, we classify it as a part of the fields. Finally, we obtain the field codes in the related class.

\noindent\textbf{\textit{Warning-Aware Slicing} \textbf{(\ding{175} and \ding{176} in~\Cref{framework})}}: 
Although we have pinpointed the exact function where the warning was issued, the data fed into our model might be noisy, e.g., the data dependency does not influence the specific bugs, or arbitrary names of local variables, hindering its ability to concentrate on the specific buggy lines and their directly related fragments. To mitigate this, we introduce a warning-aware slicing on the function. The slicing aims to (1) isolate all relevant dependencies, thereby ensuring the model's focus remains strictly on the essential elements related to the bug; (2) restructure the program and unify variable names. 

Specifically, we set the lines that the warning reported as the slice criteria, and do the backward and forward slice according to the slice criteria. We first use Soot~\cite{soot} as the analysis framework and construct the program dependency graph (PDG) of the project. For the backward slice, we focus on the variable read in the slice criteria and find all nodes in the PDG that are directly or indirectly dependent on the specified program point starting from the slice criteria. As for the forward slice, we focus on the variable written in the slice criteria and still find all nodes that are directly or indirectly dependent. Note that our slices are confined to the function's scope, as interactions between the function and class are intricate. These interactions could involve semantics associated with other warnings, thereby leading to the potential misguidance of our model. Finally, we obtain the sliced function code originally in the form of the intermediate representation (IR) in Soot~\cite{soot}, i.e., Jimple. Jimple is a typed 3-address IR. It provides a more standardized and simplified view of observing program behavior, which can assist the model in understanding the semantics of complex statements (e.g., multiple calls) and excluding the interference from the semantics of some local variables. 

\begin{figure}
\begin{lstlisting}[language=java]
$stack61 = virtualinvoke data.<java.lang.String: byte[] getBytes()>()
$stack62 = lengthof $stack61
$stack63 = staticinvoke <java.lang.Integer: java.lang.String toString(int)>($stack62)
virtualinvoke connection.<java.net.HttpURLConnection: void setRequestProperty(java.lang.String,java.lang.String)>("Content-Length", $stack63)
\end{lstlisting}
\caption{A simplified example of Jimple representation of Java code converting from ~\Cref{code1}.}
\label{code2}
\end{figure}

As shown in the ~\Cref{code2}, the code snippet is a Jimple representation of the Java code Line 5 in ~\Cref{code1}. The original Java code has been converted to three typed 3-address statements. The statements break down complex expressions and statements into simpler ones that perform a single operation and have at most one side effect. For example, the original code Line 5, involves many assignments ((e.g., \texttt{lengthof} corresponds to the \texttt{.length} property used in the original Java code) and function invoke (e.g., \texttt{getBytes}, \texttt{toString}, and \texttt{setRequestProperty}). The Jimple code divided the code in Line 5 into several different statements. In addition, the intermediate variables are simplified (e.g., $stack61$). Hence, by unifying the local variables, splitting the complex semantics, Jimple representation can guide the model more effectively to locate the bugs from the given function and messages. Again, if there is no corresponding function, we use the entire class's Jimple as a filler.

\subsubsection{Warning Information Extraction \textnormal{(\ding{177} in~\Cref{framework})}}
Besides the code context we mentioned, some warning information needs to be provided to the model. First of all, we collect the MESSAGEs of each warning, which include the details of the potential bugs and can serve as a filter for the model to identify the buggy semantics. Then, we gather the RULE and CATEGORY attributes for the model to distinguish different types of potential bugs. Finally, we also collect the RANK and CONFIDENCE of the warning for the model to judge the risk of potential bugs and possible concerns of the ASATs. The details of the warning attributes used are shown in ~\Cref{attr}.
\begin{table}[]
\caption{The details of the warning information in SpotBugs.}
\scalebox{0.90}{
\begin{tabular}{ll}
\toprule
\textbf{Information} & \textbf{Description}                                                 \\ \midrule
MESSAGE    & Error message generated by SpotBugs for warning description. \\ \midrule
RULE       & The scanning rule for triggering the warning.                 \\ \midrule
CATEGORY   & The category of the rule belongs to (e.g., CORRECTNESS).      \\ \midrule
RANK      & A specified bug severity rank of warning.                     \\ \midrule
CONFIDENCE & A particular bug confidence of warning reported.             \\ \bottomrule
\end{tabular}}
\label{attr}
\end{table}
\subsubsection{Attentional Encoder Model \textnormal{(\ding{178} in~\Cref{framework})}}
While Transformer-based architectures and LLMs have demonstrated significant success in a variety of code and natural language understanding tasks, there are important reasons for employing an LSTM-based model in our context. First, our task requires encoding multi-dimensional information for a large number of warnings in real-world projects. Utilizing LLMs or Transformer-based models for every warning would be computationally expensive and inefficient, especially when the warning volume is high. Second, most pretrained Transformer models impose strict input length limits (typically 512 tokens), which can lead to the loss of critical context when encoding larger code regions or multiple warnings. In contrast, LSTMs offer both efficient computation and the ability to flexibly handle longer or variable-length sequences, making them a practical and scalable choice for fine-grained bug-sensitive warning identification.

While modern LLMs and extended-context embedding models (e.g., Qwen, GPT-4) support extensive input sequences (8k-32k+ tokens), applying them to the domain of static analysis creates a scalability bottleneck. ASATs frequently generate tens of thousands of warnings; processing such a massive volume with heavy-weight LLMs incurs prohibitive computational costs and latency, making them unsuitable for real-time Continuous Integration (CI) pipelines.

Hence, our \revise{approach} applies a relatively lightweight model. This enables us to explore dedicated strategies for capturing more precise semantics. In particular, we design an attention encoder model, including five embedding layers, four LSTM models, two max-pooling layers, two cross-attention modules and two fully connected layers. The five embedding layers and four LSTM models separately encode the function code, the field code, the sliced IR, the warning message, and SpotBugs warning attributes (i.e.,  RULE, CATEGORY, RANK, and CONFIDENCE features). This rationale for choosing LSTM is supported by prior research~\cite{lstmuse1}~\cite{lstmuse2}~\cite{lstmuse3}, which has demonstrated the effectiveness in similar contexts. Given that warning messages typically reflect the semantics embedded within the code, we independently perform cross-attention between the function code and its corresponding warning message to enhance semantic alignment. After encoding the field code, the sliced IR, and warning attributes through max pooling and full connection. Max pooling was chosen because it effectively extracts the most prominent features from local regions of the encoded sequences—filtering out noise and reducing dimensionality—to yield robust representations for further processing. The fully connected layer is utilized to integrate the diverse contextual cues into a unified high-level representation, enabling the model to capture complex interactions and accurately classify warnings as bug-sensitive or bug-insensitive. Finally, we obtain 5 semantic vectors. 

Formally, let $f$ denote the sequence of words $\{s_1,s_2,...,s_n\}$ in a specific function, and an embedding layer is applied to initialize the sequence by vectors:
\begin{equation}
    f_i=W_e^\top s_i, i \in [1,n],
\end{equation}
where n is the length of the sequence (i.e function $f$), $W_e$ is the embedding matrix. Then, we employ a Long Short-Term Memory (LSTM)~\cite{LSTM} to encode the sequence of vectors $F = \{f_1,f_2,...,f_n\}$ into hidden states $H_f = \{h_1,h_2,...,h_n$\}. To capture the semantics before and after the current position, we further use a Bidirectional LSTM (Bi-LSTM)~\cite{biLSTM} on it. In this step, we formalized the result is
\begin{equation}
    h_j = Bi\text{-}LSTM (f_j, h_{j-1})
\end{equation}
Similarly,  we obtain the hidden states of field code, sliced IR, and warning message by three new Bi-LSTMs. For ease of illustration, we mark them as $H_{fc}$, $H_{J}$ and $H_{wm}$ respectively.

After encoding all the code and message, we design a cross-attention mechanism on the vector representations of function code and warning message. The two cross-attention blocks in~\Cref{framework} (center) implement this bidirectional conditioning between the function-code sequence and the warning-message sequence. Intuitively, cross-attention allows the model to answer questions such as ``given this warning message, which parts of the method are most relevant?'' and ``given this method, which words in the warning message matter most?''. To this end, we compute an attention-weighted summary of the warning message conditioned on the encoded method, and symmetrically an attention-weighted summary of the method conditioned on the warning message. These two summaries capture the most informative code-warning interactions for classification.

Formally, given the hidden states of function code are $H_f = \{h_{f_1}, h_{f_2},...,h_{f_n}\}$, we first calculate the query vector $q_f$ by  max pooling $H_f$ as follows:
\begin{equation}
q_f = \text{maxpooling}(H_f) 
\end{equation}
The context vector \(V_f\) is computed using the following formula:
\begin{equation}
V_f = \sum_{i=1}^{n} \alpha_{i} \cdot h_{m_i} 
\end{equation}
where \(n\) is the number of hidden states in the function code, and \(\alpha_{i}\) is the attention weight for each hidden state \(h_{m_i}\), calculated as follows:
\begin{equation}
     \alpha_{i} = \frac{\exp(q_s \cdot h_{m_i})}{\sum_{j=1}^{n} \exp(q_s \cdot h_{m_j})} 
\end{equation}
The context vector \(V_f\) is the weighted sum of the hidden states based on their attention weights, allowing the model to focus on the most relevant semantics of the warning message. For simplicity, we notate the process as $V_f=Attention(q_f, H_m)$. In this way, we depict the relationships between the whole semantics of the sliced IR code and the words in the warning message.

Following the process, we can obtain the context vector of the message $V_m$, which captures the relationships within the sliced function through a query vector $q_m$. That is, $V_m=Attention(q_m, H_f)$. By computing the attention weights in a mutual manner, we can better acquire the correlations between the sliced code and the warning message.

For the field code vector, $H_{fc}$, we apply a max pooling on it, to achieve a semantics vector by 
$V_{fc}=maxpooling(h_{fc})$. Similarly, for sliced IR, we obtain a semantics vector $V_J$ by max pooling. Besides,
to leverage other attributes of warnings, we encode these attributes separately as vectors. Suppose that the \textit{t}-th attributes is $a_t$,  where $W_{e_t}$ is the weight of the attributes $t$, we embed it into $x_t = W_{e_t}^\top a_t, t \in [1,4] $ and then apply a fully connected layer on it, i.e., $V_{a_t}=W_a^\top x_t + b_t$. These encoded attributes are then combined into an integrated feature vector, denoted as $V_a$. 
As a result, we obtain 5 vectors for the semantics from different aspects.

\subsubsection{Feature Fusion \textnormal{(\ding{179} in~\Cref{framework})}}
Based on the above calculations, we have obtained five semantic vectors: field code vector $V_{fc}$,  function code vector $V_{f}$, sliced IR representation vector $V_J$, warning message vector $V_{wm}$, and other warning attributes vector $V_a$. Further, we need to integrate these four vectors to obtain a single vector by concatenating the vector. The calculation is as follows:
\begin{equation}
    V= V_{fc} \oplus V_{f} \oplus V_{J} \oplus V_{wm} \oplus V_a
\end{equation}
Finally, we employ a fully connected layer on the vector $V$ to get the final vector $V_l$.

\subsubsection{Focal Loss}
Since ASAT will explode with a large number of warnings, the data we collected also face a situation where bug-sensitive warnings and bug-insensitive warnings samples are extremely imbalanced (i.e., close to 1:15). Therefore, during the training process, we used focal loss~\cite{lin2018focal} to handle this situation. The Focal Loss function handles the issue of class imbalance in training exercises such as object detection. It modifies the cross entropy loss with a modulating term to concentrate on challenging instances of misclassification. The Focal Loss function is essentially a cross-entropy loss that is scaled dynamically, where the scaling component reduces to zero as the certainty in the accurate class rises. Hence, training our \revise{approach} is to minimize the loss function:
\begin{equation}
    FL(p_t)= -\alpha_t(1-p_t)^\gamma log(p_t)
\end{equation}
\begin{equation}
    p_t=exp(\frac{1}{N}\sum_i -[y_i log(p_i)+(1-y_i) log(1-p_i)])
\end{equation}
where $\alpha$ and $\gamma$ is the tunable focusing parameter, $y_i$ is the label of $i$, $p_i$ is the probability of predicting a bug-sensitive warning. To train the model, we apply the sigmoid activation function as the last linear layer for warning identification, and obtain the vector $L$, which represents the possibility that the corresponding warning is bug-sensitive. 
After optimizing all the parameters, the models are trained. They output probabilities $L$ for different labels. Therefore, we obtain the prediction using the following equation:
\begin{equation}
    \text{Prediction}=
    \begin{cases}
    \text{bug-sensitive}, & L>\theta\\
    \text{bug-insensitive}, & L\leq\theta
    \end{cases}
\end{equation}
where $\theta$ is the threshold.

%% file: sections/4-evaluation.tex
\section{Evaluation}
\subsection{Dataset Evaluation}
Since existing datasets are not applicable to our fine-grained warning identification task, we collected \ourdataset with 280,273 warnings according to our \revise{approach}. However, the quality of the dataset may still affect the validity of our \revise{approach}. Although we have employed a keyword filter and warning comparison approach to avoid invalid warnings, it is possible to introduce some warnings with incorrect labels. Hence, we conduct a manual evaluation of the dataset. We used \revise{stratified random sampling~\cite{aoyama1954study}} to ensure representation across different warning types. 
To determine the appropriate sample size, we applied the formula for estimating proportions with a 95\% confidence level and a 5\% margin of error. Additionally, considering the class imbalance in \ourdataset, we allocated samples proportionally to each class to accurately reflect their distribution. Finally, we sampled 27 bug-sensitive warnings and 357 bug-insensitive warnings from \ourdataset to evaluate the labeling accuracy.

Then, three experienced senior developers from our industry partner, each with frequent experience using SpotBugs, independently evaluated the sampled warnings. For every sampled instance, evaluators were provided with: (\textit{i}) the SpotBugs warning message and attributes, (\textit{ii}) the buggy and fixed versions of the affected file, and (\textit{iii}) the corresponding bug-fix diff. The 
developers were instructed to determine the correctness of the assigned labels based on the following criteria:
\revise{(\textit{i})} For bug-sensitive warnings, the developers checked whether the warning's location overlapped with code changes that directly addressed the defect described by the bug-fix commit, according to our formal definition in \Cref{problem_definition}. \revise{(\textit{ii})} For bug-insensitive warnings, they assessed whether the warning was unrelated to the bug-fix (e.g., disappeared due to unrelated refactoring or remained unchanged) and whether it might nonetheless indicate another defect.

We first asked the three senior developers to individually assess the sampled data.
After the individual assessments \revise{(the Kappa score in this process is 0.728, indicating substantial agreement~\cite{landis1977application})}, the three evaluators discussed all cases and resolved disagreements by consensus, producing a final ground-truth label for each sampled warning. Ultimately, the three evaluators reported that among the 27 bug-sensitive warnings, 2 were mislabeled, and among the 357 bug-insensitive warnings, 13 were mislabeled. In conclusion, \ourdataset has an accuracy rate of 92.60\% on bug-sensitive warnings and 96.36\% on bug-insensitive warnings with a confidence of 95\%.

\subsection{Experiment Setup}
We carried out the experiments on \ourdataset originally consisting of 280,273 warnings. For the purposes of training, validation, and testing, we divided \ourdataset in an 8:1:1 ratio. Additionally, we applied our trained model to four open-source projects to evaluate the performance of \ourtool in real-world scenarios.

We implement \ourtool based on pytorch 1.8.0. In the training process, we set the vocabulary size to 100,000, the embedding size to 512, and the hidden state size of LSTM to 512. We set the batch size to 64 and adopt the Adam~\cite{kingma2017adam} as the optimizer. The learning rate is 5$e$-5 with a decay when the F1-score for the bug-sensitive samples has stopped improving. The hyperparameters of focal loss $\alpha$ and $\gamma$ are set to 0.05, 2, respectively. All aforementioned hyper-parameters are chosen from a range of alternatives, based on their performance on the validation set. All the experiments are conducted on one Ubuntu 18.04 server with 48-core CPU, 256GB RAM, and two NVIDIA RTX 3090 GPUs with 24GB of memory.
\subsection{Baselines}
\ourtool represents the first of its kind in identifying ASAT warnings at a fine-grained warning level based on deep learning models. To demonstrate its effectiveness, we conducted a comparative analysis with existing representative learning-based \revise{approaches}. These \revise{approaches} include: ``GHOST2~\cite{golden,Yang,kang,yedida2023find}'',   ``DeepInferEnhance~\cite{Kharkar}'',``FRUGAL~\cite{FRUGAL}'', and ``LLM4SA~\cite{Wen}''. Additionally, we also implemented four straightforward solutions including ``GraphCodeBERT model~\cite{graphcodebert}'', ``transformer model~\cite{vaswani2017attention}'', ``LSTM model~\cite{LSTM,biLSTM}'', and ``N-gram model~\cite{brown1992class}''. In our comparisons, we adopted the default settings for the above \revise{approaches} unless specified otherwise. 

\subsubsection{SOTA \revise{Approaches}}
\begin{description}[style=unboxed,leftmargin=0cm]
    \item[GHOST2:] 
    Wang \etal and Yang \etal conducted a study on existing features to predict actionable warnings and identified key attributes as ``23 golden features''~\cite{golden,Yang}. The features include 8 categories, involving file, history, code, and warnings. Based on this, Kang and Yedida \etal continued the research on golden features and proposed GHOST2 to achieve the best performance~\cite{kang,yedida2023find}. Hence, we followed the same settings of Kang and Yedida \etal \revise{approaches} to replicate this.
    \item[FRUGAL:] 
   In addition to the aforementioned supervised learning \revise{approaches}, there is a semi-supervised learning \revise{approach}, FRUGAL, for actionable warning identification. FRUGAL, was introduced by Tu \etal~\cite{FRUGAL}, and achieved nearly the same result with supervised learning in actionable warning identification. For this baseline, we use the code provided in their official repository.
    \item[DeepInferEnhance:]
    This approach~\cite{Kharkar}, introduced by Kharkar \etal,  employs a pre-trained model CodeBERTa to determine if code functions flagged by warnings contain bugs. We specifically chose CodeBERT ~\cite{codebert} for its superior performance, and establish it as a baseline for our study. In addition, as our work takes SpotBugs as input and carries out fine-grained warning prediction, we also made some modifications to the DeepInferEnhance model. We input both the warning message and code simultaneously and used a fully connected layer to fuse these two aspects, making it our baseline.
    \item[LLM4SA:]
     Wen \etal~\cite{Wen} conducted a study using LLMs to investigate LLMs' ability to identify ASAT warnings. They developed the LLM4SA framework and evaluated it in three C++ ASATs. For this baseline, we followed the same setting in the paper and used the GPT-3.5 version of LLM4SA which outperformed LLama-2 in the paper. Additionally, we also implemented LLM4SA using DeepSeek-V3~\cite{deepseekv3} as another baseline, which is an open-source model with lower cost.
\end{description}

\subsubsection{Straightforward Solutions}
\begin{description}[style=unboxed,leftmargin=0cm]
    \item[GraphCodeBERT Model:] 
    In addition to the aforementioned CodeBERT model in DeepInferEnhance, we also used another pre-trained model GraphCodeBERT~\cite{graphcodebert}. This model incorporates data flow into the foundation of CodeBERT, resulting in superior semantic understanding compared to CodeBERT. There are some studies that use GraphCodeBERT to detect bugs and vulnerabilities~\cite{wu2021peculiar,ni2022defect}. For this baseline, we used the same settings as CodeBERT. The input lengths of the code and the message are 512 and 32, respectively.
    \item[Transformer Model]
    We additionally employ a common transformer~\cite{vaswani2017attention} to replace the pre-trained model, which can effectively extract semantic information from code. There are also some studies~\cite{allamanis2021self,pornprasit2022deeplinedp} to predict bugs and vulnerabilities based on it. Samely, we concatenate the code and message together and use a fully connected layer to obtain the label. The input size of the code and message are 512 and 32. The hidden dimension of the transformer is 1024. The numbers of heads and encoders are 8 and 6, respectively. 
    \item[LSTM Model:] 
    In this model, we use LSTM~\cite{LSTM,biLSTM} as the encoder to replace the transformer. We encode the code and message separately and concatenate them. The hidden dimension of LSTM is 1024, and the n layers of LSTM is 1. Based on this, we apply a fully connected layer to obtain the label.
    \item[N-gram Model:]
    Lastly, we use an N-gram\cite{brown1992class} model as a baseline. N-gram are widely used in natural language processing and computational linguistics. We concatenate the code and message together as input to verify the validity of the warning (N=3). We built a vocabulary using TF-IDF~\cite{ramos2003using} and applied a linear SVM as the machine learning \revise{approach}, ultimately obtaining our baseline.
\end{description}

To ensure a fair comparison, we adopted different training strategies tailored to the model architectures. \ourtool, Transformer Model, and LSTM model were trained from scratch on the BSWarnings dataset, initialized with random weights. For the pre-trained Transformer-based baselines (CodeBERT and GraphCodeBERT), we initialized them with their public pre-trained checkpoints and performed full fine-tuning on the BSWarnings dataset to adapt their representations to the target task.

\subsection{Results and Discussion}
We present the results of our experiments and delve into the analysis by answering the following research questions (RQs).

\subsubsection{RQ1: How effective is \ourtool compared to the baseline models?}

\begin{table}[]
\caption{Results comparison with different baselines for warning identification (DeepInferEnhanceWM refers to DeepInferEnhance with warning messages, LLM4SA\_GPT refers to LLM4SA using GPT-3.5-turbo model, and LLM4SA\_DeepSeek refers to LLM4SA using DeepSeek-V3 model).}

\scalebox{1}{
\begin{tabular}{lccc}
\toprule
Methods & Precision (\%) & Recall (\%)    & F1-Score (\%) \\ \midrule

GHOST2  & 51.47          & 51.04          & 51.25 \\ 
DeepInferEnhance & 49.59          & 44.89          & 47.12          \\ 
DeepInferEnhance\_WM  & 59.51          & 47.86          & 53.05     \\ 
FRUGAL       & 5.62           & \textbf{85.53} & 10.36           \\ 
LLM4SA\_GPT &7.21&	83.47&	13.28 \\
LLM4SA\_DeepSeek & 7.03 & 32.25 & 11.55 \\ \hdashline
GraphCodeBERT  &59.30 &47.00& 52.44\\ 
Transformer & 51.03 &51.99& 51.51 \\ 
LSTM  &59.45 &49.28& 53.89\\ 
N-grams  &66.49 &39.96& 49.92\\ \midrule

\ourtool   & \textbf{75.52} & 60.31          & \textbf{67.06}  \\ \bottomrule
\end{tabular}}
\label{result}
\end{table}

~\Cref{result} presents a comparative analysis of \ourtool and various baseline models in terms of their ability to identify bug-sensitive warnings using metrics precision, recall, and F1-score. \revise{The dashed line separates the SOTA approaches and straightforward solutions.}
The highest values across all columns are highlighted in bold. 
As depicted in ~\Cref{result}, \ourtool effectively reduces the cost of manually inspecting warnings. It can identify 60.31\% of bug-sensitive warnings with 75.52\% precision. Overall, \ourtool performs with 67.06\% F1-score in the warning identification at a finer granularity.

Compared to other \revise{approaches}, \ourtool generally outperforms the eight baseline models. The ``GHOST2'' \revise{approach}, which previously performed well, appears to be less effective when dealing with more fine-grained bug-sensitive warnings, with a 51.25\% F1-score. The primary reason is that fine-grained bug warning detection demands the capture of extensive semantic information. However, the \revise{approach} in question relies on a set of fixed features, which are insufficient to cover the diversity and complexity of bug semantics. The manually extracted semantics may provide identical features in certain scenarios that are challenging to distinguish, thereby making it difficult for the classifier to differentiate them and resulting in a notably low recall and precision. ~\Cref{code1} is a typical example where ``GHOST2'' performs failed in the warning identification. SpotBugs reported a warning in the function \texttt{sendRequest} of lines 5 and 10 with the same warning type. Hence, the warning characteristic features and code features can not distinguish them. Besides, the codes are both introduced in the first commit of the project, the features of warning history also failed in finding differing aspects of them. Consequently,  the classifier recognizes the two warnings are the same warnings and fails to identify these warnings.    

Despite DeepInferEnhance undergoing pre-training on a large code corpus, it fails to meet the characteristics of this challenging task. A notable limitation arises when the input lacks a warning message; the original model struggles to accurately differentiate between various warnings within the same function. This leads to a modest F1-score of 47.12\%. In contrast, incorporating warning messages into the analysis improves the F1-score by 5.93\%. As for another pre-trained model, GraphCodeBERT, which have shown better performance than CodeBERT in other tasks, also struggles to accurately identify bug-sensitive warnings, with only 52.44\% F1-score.  While these pre-trained models are adept at understanding the general behaviors of code, they neither deal with the noise (e.g., some unrelated variable), nor depict the correlation between the code and the warning information. Additionally, these models break the identifiers down into sub-tokens, which makes it hard to capture the dependencies between those elements. This result is consistent with previous studies that acknowledged the limited advantage of pre-trained models in defect prediction~\cite{plbart}.

FRUGAL, as a semi-supervised learning \revise{approach}, previously demonstrated comparable performance to supervised feature-based \revise{approaches} in actionable warning identification. However, its effectiveness significantly diminishes in the bug-sensitive warning identification, achieving only a 10.36\% F1-score. 
One reason is the inherent challenge FRUGAL faces with extremely imbalanced datasets. In such scenarios, models may develop a bias towards the class with higher weight due to its more abundant information, leading to subpar predictions for the minority class ~\cite{semi1,semi2,semi3}. In our experiments, to improve the F1-score for bug-sensitive warning identification, the model tended to over-predict bug-sensitive warnings. This approach increased recall but adversely affected precision, this is why FRUGAL achieved a high recall of 85.53\% but a very low precision of 5.62\%. It tended to classify most warnings as bug-sensitive warnings which is in vain for the purpose of reducing manual inspection costs.

Additionally, this strategy also resulted in a decreased recall for bug-insensitive warnings. Another contributing factor to FRUGAL's under-performance in finer granularity tasks is the potential inappropriateness of its original features for the nuanced requirements of bug-sensitive warning identification.

LLM4SA, which leverages a language model-based approach, demonstrates potential in identifying bug warnings within static analysis results. However, it falls short in accurately detecting bug-sensitive warnings, achieving an F1-score of only 13.28\%. Similiar to FRUGAL, LLM4SA also tends to classify most warnings as bug-sensitive warnings, leading to a high recall of 83.47\% but a very low precision of 7.21\%, which is ineffective for reducing manual inspection costs.
 It is also important to note the considerable cost associated with this approach: our experiments with GPT-3.5-turbo-0125 incurred expenses exceeding \$100 per run. Even when using open-source models such as DeepSeek-V3, which also achieved a lower F1-score of 11.55\%, substantial computational resources were required (e.g., 8× H100 GPUs). In contrast, DeepFWI not only outperforms \revise{the LLM-based approach} in terms of F1-score, but can also be efficiently executed on a single consumer-grade GPU (e.g., NVIDIA 3090), resulting in significantly lower hardware and financial costs. This comparison highlights the practical advantages of DeepFWI for scalable and cost-effective deployment.

Additionally, the limited effectiveness of LLM4SA can largely be attributed to the ``hallucination problem~\cite{rawte2023survey},'' a common challenge with language models where they generate plausible but incorrect or irrelevant responses. This issue is exacerbated in the complex and nuanced realm of bug semantics found in real-world datasets~\cite{Purba,fang2024large}. LLM4SA solely relies on inputting code contexts and prompts to identify warnings. Due to the complexity of bug semantics in the real world, LLMs without fine-tuning struggle to capture these intricate bug semantics. Therefore, it fails to identify bug-sensitive warnings in real-world scenarios. This aligns with the transition from the artificial Juliet Test Suite to real-world datasets results in the real-world projects in~\cite{Wen}. Additionally, LLMs are significantly influenced by prompts. In Wen \etal.'s \revise{approach}, prompts tend to guide the model to perceive warnings as true, leading LLM4SA to classify most warnings as real bugs. Similarly, existing studies have shown that LLM models can produce erroneous results in bug detection, necessitating further verification through static analysis (SA) technologies ~\cite{sun2024gptscan}. 

Regarding basic alternatives, their performance still falls short of adequacy. The transformer model, lacking pre-training, faces challenges in  learning code semantics and interpreting warning messages. This difficulty in training resulted in a suboptimal performance, achieving only a 51.51\% F1-score. The n-gram \revise{approach}, another baseline, exhibits average performance, yet a little lower than ``GHOST2''. This suggests that while the n-gram model can capture related patterns, its performance is hampered by noise and inability to learn deep semantics. We also find that as the base component of FineWAVE, the LSTM model out performed the pre-trained models. This can be partly attributed to its strict orderliness of encoding, which makes it easier to learn the naturalness of source code, thereby acquiring 
more sequential dependencies. Therefore, their performance was not as good as that of the LSTM.

Additionally, in order to further present the improvement of the effectiveness of \ourtool, we also conducted a paired t-test on the F1-score of \ourtool and the baseline models. The results show that \ourtool significantly outperforms all the baseline models with a p-value of 0.0049. This indicates that the improvement of \ourtool is statistically significant. Furthermore, the improvement of \ourtool is quite outperforming compared to some other works in the same field of software engineering~\cite{Kharkar,Wen,Zhang_2023_Detecting,Cheng_2021_DeepWukong,Li_2022_VulDeeLocator}.
In conclusion, \ourtool considers more elements (including field code and warnings), integrates the function slicing to obtain rich while precise context and takes into account the interaction between warning messages and code. These enhancements allow \ourtool to outperform the baseline models, underscoring its effectiveness in identifying warnings.

\noindent\fbox{
	\parbox{0.95\linewidth}{
		\textbf{Answer to RQ1:} \textit{Compared with the ten baseline models, \ourtool demonstrated superior performance, achieving improvements of at least 9.03\% in precision,  and 13.17\% in F1-score. These results underscore the effectiveness of \ourtool in identifying bug-sensitive warnings.}}
}

\smallskip

\subsubsection{RQ2: How effective are the individual components of \ourtool?} 
\label{rq2}
In this RQ, we explore the impact of the key components of \ourtool on its overall effectiveness. Our methodology primarily incorporates four components: warning-aware slicing, field code, cross-attention mechanism, and focal loss.
To evaluate the significance of the individual components, we conducted an ablation study of \ourtool by removing each component and observing the consequent effects on performance. Moreover, to further assess the impact of our design choices, we experimented by substituting the LSTM in our encoder with a transformer architecture.

\begin{table}[]
\caption{Effectiveness of the main components of \ourtool.}
\scalebox{0.90}{
\begin{tabular}{lccccccccc}
\toprule
\textbf{Components}             & \multicolumn{1}{l}{Precision (\%)} & \multicolumn{1}{l}{Recall (\%)} & \multicolumn{1}{l}{F1-score (\%)} \\ \toprule

W/O Field Code                  & 71.08                              & 57.66                           & 63.67                             \\ \midrule
W/O Warning-Aware Slice         & 70.88                              & 61.58                           & 65.90                             \\ \midrule
W/O Cross-Attention             & 66.51                              & \textbf{64.38}                  & 65.43                              \\ \midrule
W/O Function Code               & 73.56                              & 59.34                           & 65.69                     \\ \midrule
W/O RULE                        & 71.05                              & 62.19                           & 66.32                      \\ \midrule
W/O CATEGORY                    & 72.71                              & 61.68                           & 66.74                       \\ \midrule
W/O RANK                        & 71.16                              & 62.04                           & 66.29                      \\ \midrule
W/O CONFIDENCE                  & 70.99                              & 62.39                           & 66.41                     \\ \midrule
W/O Focal Loss                  & 70.04                              & 61.63                           & 65.57                            \\ \midrule
LSTM\textrightarrow Transformer & 67.30                              & 59.39                           & 63.10                            \\ \midrule
\ourtool                        & \textbf{75.52}                     & 60.31                           & \textbf{67.06}                   \\ \bottomrule
\end{tabular}}
\label{remove}
\end{table}

The impact of each component of \ourtool on its effectiveness is detailed in ~\Cref{remove}. Our first step was to remove the Field Code component, which resulted in a 3.39\% decrease in the F1-score. This decline can be attributed to the diminished ability of the model to fully understand certain function code statements in the absence of the class variable segment provided by the Field Code. Consequently, this led to the model missing warnings that may contain bugs, resulting in false positives and inaccurate identification.

Upon removing the slicing component, a decline in \ourtool's effectiveness was observed, with a 1.16\% decrease in the F1-score. Without this component, the model can encounter some noises (e.g. unrelated data flow in the same function), which adversely affects the tool's focus on locating the bug information. Consequently, the precision of warning identification decreases by a large margin. Similarly, without the Cross-Attention component, there was a 1.63\% decline in the F1-score, and notably, a substantial 9.01\% drop in precision. As discussed earlier, the Cross-Attention mechanism is indeed important for facilitating the mutual integration of warning messages and code. The results indicate that it is essential to consider precise semantics for fine-grained warning identification. 

Furthermore, we evaluated the impact of excluding the Function Code component, which is responsible for providing the related information of the warning for the model. Removing this component led to a decrease in the F1-score by 1.37\%. Specifically, removing Function Code resulted in a decrease in precision by 1.96\%. The results suggest that Function Code provides an overall context, allowing the model to build a preliminary understanding of the program’s general structure and semantics. In \ourtool, while the function code sets the stage with broad context, the sliced IR zooms in on the critical areas that ultimately impact bug detection, thereby contributing to increased precision.

Additionally, we assessed the impact of excluding the RULE, CATEGORY, RANK, and CONFIDENCE components, which is the other warning information provided to the model. Removing these components led to a decrease in the F1-score by 0.74\%, 0.32\%, 0.77\%, and 0.65\%, respectively. Specifically, replacing these components with the warning message resulted in a decrease in precision by 4.47\%, 2.81\%, 4.36\%, and 4.53\%, respectively. The results suggest that these components provide additional information to the model, enabling it to build a more comprehensive understanding of the warning and focus on the potential bugs, thereby contributing to increased precision.

We also assessed the impact of excluding the Focal Loss component, which was employed to manage the significant imbalance in \ourdataset.
Removing this component led to a modest decrease in the F1-score by 1.49\%. Specifically, replacing Focal Loss with Cross-Entropy Loss resulted in a decrease in precision by 5.48\%.
The results suggest that Focal Loss's emphasis on less represented samples enhances the model's ability to gather more bug-related information, thereby contributing to increased precision.

Finally, when we substituted the LSTM encoder in \ourtool with a Transformer model, contrary to expectations, there was a decrease in performance, dropping to 63.10\% while we expect the self-attention mechanism of the Transformer to enhance \ourtool's performance. The Transformer model processes tokens differently from LSTM, and this fundamental variance in semantic extraction could be the reason behind the diminished performance. Specifically, the self-attention mechanism of Transformers may focus on too many irrelevant dependencies, which makes LSTM a more suitable choice for our specific task.

\smallskip
\noindent\fbox{
	\parbox{0.95\linewidth}{
		\textbf{Answer to RQ2:} \textit{
The effectiveness of each component of \ourtool varies slightly, impacting performance by 0.32\% to 3.96\%. Field Code improves recall by broadening the context, while Warning-Aware Slicing and Cross-Attention enhance precision by focusing on more accurate semantics. Focal Loss, Function Code, the RULE, CATEGORY, RANK, and CONFIDENCE components slightly increases the F1-score and precision by better semantic extraction.
	}}
}
\smallskip

\subsubsection{RQ3: What is the effectiveness of \ourtool across different warning categories?}
In this RQ, we analyze and discuss the effectiveness of \ourtool in various categories of SpotBugs. Our classification comes from SpotBugs' categorization of its detection rules. There are a total of 10 types of categories in SpotBugs, i.e., \textit{Correctness} (apparent coding mistakes), \textit{Experimental} (experimental bug pattern), \textit{Internationalization}, \textit{Malicious code}, \textit{Multithreaded correctness}, \textit{Bogus random noise}, \textit{Performance}, \textit{Security}, and \textit{Bad Practice}. 
Because the \textit{Bogus Random Noise} category intended to be useful as a control in data mining experiments, not in finding actual bugs in software, we excluded it from our analysis.  

As shown in ~\Cref{category}, \ourtool achieves high performance in the \textit{Bad Practice}, \textit{Correctness}, and \textit{Performance} categories, with overall F1-scores about 70\%. \textit{Bad Practice} warnings, often involving simpler bug patterns related to variable issues, are most effectively detected due to the model's focus on field code and variable-related semantics. In the \textit{Correctness} category, which includes a variety of potential coding errors, our tool performs well, though some complex bugs needing more information may not be detected. The \textit{Performance} category, similar to \textit{Bad Practices}, shows good results, especially in identifying issues like unused code and inefficient loops. However, our tool might not always catch performance issues that are indirectly linked to bugs.

For \textit{Security} category, \ourtool performs worst on it, only achieving an F1-score of 36.36\%. It is worth noting that there are very few bug-sensitive warnings in this category. \ourdataset contains a limited number of bug-sensitive warnings in the Security category, totaling only 63 instances. This scarcity of examples for the model to learn from could be a contributing factor to its low performance in validating issues within this specific category. This phenomenon also aligns with the result in the Fluffy on the classification of unexpected path-traversal flow~\cite{Chow}. Besides, accurately determining if these rules align with specific security issues in the code demands consideration of external factors, like inter-procedure dependencies and the execution environment.

For the other five categories, \ourtool's performance hovers around a 60\% F1-score, consistent with its overall effectiveness. Specifically, \textit{Multithreaded Correctness} and \textit{Experimental} exhibit similar levels of precision and recall which indicates that \ourtool presents a balanced ability on them due to its components. The other three categories, however, show a stronger performance in precision. This pattern suggests that while \ourtool is adept at the detailed analysis of code and warnings, it sometimes misses nuances related to specific bug dependencies. The results across these categories reflect the tool's capability in handling a variety of code issues, though it also highlights areas for potential refinement in targeting bug-specific dependencies.

\begin{table}[]
\caption{Effectiveness of \ourtool on each category in SpotBugs.}
\scalebox{0.90}{
\begin{tabular}{lccc}
\toprule
\textbf{Categories}     & \multicolumn{1}{l}{Precision (\%)} & \multicolumn{1}{l}{Recall (\%)} & \multicolumn{1}{l}{F1-score (\%)} \\ \toprule
\textit{Bad Practice}            & \textbf{83.46}                    & \textbf{69.00}                 & \textbf{75.54}                   \\ 
\textit{Performance}             & 76.51                             & 64.96                         & 68.47                            \\
\textit{Correctness}             & 68.68                             & 67.57                         & 68.12                            \\ \hdashline
\textit{Malicious code}          & 76.15                             & 57.70                          & 65.65                            \\
\textit{Dodgy code}              & 75.34                           & 56.85                          & 64.80                           \\
\textit{Experimental}            & 66.67                             & 62.50                          & 64.52                            \\
\textit{Internationalization}    & 71.28                             & 53.17                          & 60.91                           \\
\textit{Multithreaded correctness} & 60.53                           & 57.50                        & 58.97                            \\ \hdashline
\textit{Security}                & 50.00                             & 28.57                          & 36.36                            \\

\midrule
\ourtool                & 75.52                             & 60.31                          & 67.06                            \\ \bottomrule    
\end{tabular}}
\label{category}
\end{table}

\smallskip

\noindent\fbox{
	\parbox{0.95\linewidth}{
		\textbf{Answer to RQ3:} \textit{\ourtool performs differently across various categories, with F1-scores ranging from 36.36\% to 75.54\%, precision ranging from 50.00\% to 83.46\%, and recall ranging from 28.57\% to 69.00\%. Among them, our model performs best in the Bad Practice category, and worst in the Security category mainly due to a small number of bug-sensitive warnings for it.} 
	}
}

\smallskip

\subsubsection{RQ4: Can our tool generalize when applied to real-world scenarios, and can it help developers identify new bugs?}
\label{rq4}
In this RQ, we aim to evaluate the effectiveness in real-world scenarios from the perspective of a developer. Specifically, we aim to explore the effectiveness of \ourtool in practical scenarios instead just on our datasets. We discuss how our model's generalization in the latest (unseen) versions of projects, and whether it can identify (in)valid warnings. Due to the fact that the latest version does not have a fixed version for bugs, i.e., there is no ground truth for the latest version, we cannot automatically evaluate the performance of \ourtool in the latest version. However, we can still evaluate the effectiveness of \ourtool in the latest version by manually checking the warnings. Hence, we applied \ourtool to 4 popular OSS projects which have over 1.5k stars, i.e., itext7, metadata-extractor, poi, and pdfbox.

Initially, we compiled the code and used SpotBugs to scan them, respectively.
By analyzing the SpotBugs reports, we totally obtained 7,235 warnings from 4 projects. 

Then, we applied \ourtool on these warnings to filter out the false positives (i.e. bug-insensitive warnings). 
In this step, there remained 551 warnings (7.62\%) in all projects. 
Based on these filtered warnings, we performed a manual check on each warning. 
For each warning, we not only analyze the warning's type and information to determine if the reported line is the root cause, but we also search for corresponding call issues based on the clues provided by each warning. 
This allows us to confirm whether the warning is part of the triggering chain of certain bugs. Finally, we totally found 25 bugs in the warnings validated by \ourtool. We have submitted these bugs to the relevant developers and 4 of them have been confirmed and fixed by developers. In order to explore whether there may be a potential bug in the remaining warnings, we also conducted a sample analysis of the remaining 6,684 warnings. We randomly selected 200 warnings and manually checked them. Finally, we did not find any bugs in these warnings. Hence, we indicate that the remaining warnings are most bug-insensitive warnings.
\begin{figure}
\begin{lstlisting}[language=Java]
public void nextValidToken() throws java.io.IOException {
... 
    assert n2 != null;
    type = TokenType.Obj;
    reference = Integer.parseInt(new String(n1));
    generation = Integer.parseInt(new String(n2));
    return;
... 
}
\end{lstlisting}
\caption{A simplified bug example of \texttt{nextValidToken} method in itext7.}
\label{code:bug}
\end{figure}
\noindent\textit{\textbf{Case Study:}}
As shown in the ~\Cref{code:bug}, after filtering through our tool, we obtained two warnings: ``Found reliance on default encoding in lines 5 and 6''. We manually analyzed these two warnings and conducted an analysis of the entire function. We found that line 5 and line 6 could potentially lead to problems if the byte arrays contain data that is not compatible with the default charset. If the bytes do not represent valid characters in the default charset, the resulting string could contain unexpected or incorrect characters, which could then cause \textit{Integer.parseInt()} to throw a ``NumberFormatException''. 
To avoid this potential issue, it is generally recommended to specify a charset when converting bytes to a string. We have reported this issue to the developers, and now it employs an error handling mechanism to ensure that the byte arrays are properly converted to strings using a specified charset, thereby preventing potential errors related to character encoding.

\smallskip

\noindent\fbox{
	\parbox{0.95\linewidth}{
		\textbf{Answer to RQ4:} \textit{When applying \ourtool on four OSS projects with over 1.5k stars (i.e., itext7, metadata-extractor, poi, pdfbox), \ourtool can help remove 92.38\% of the warnings with potential false alarms, and enable us to discover 25 new bugs in these projects, 4 of which have been confirmed and fixed by developers.}}
}

%% file: sections/5-threats.tex
\section{Discussion}
\subsection{Threats to Validity}
\subsubsection{External Validity.}
As detailed in \S~\ref{datacollection}, our data collection process involved identifying bug-related keywords in commit messages from GitHub's projects. However, due to potential inaccuracies in commit descriptions, some commits may not be bug fixes or may contain unrelated changes, impacting \ourdataset's quality. Furthermore, the process of collecting warnings may also involve some poor-quality warnings which are not related to the specific bug. Hence, we manually sampled 384 warnings to assess \ourdataset, finding an accuracy rate of 92.60\% for bug-sensitive warnings and 96.36\% for bug-insensitive warnings, with a 95\% confidence level.

\subsubsection{Internal Validity.}
A potential threat to internal validity refers to the generalizability across ASATs.
Following the existing research focusing on single ASATs~\cite{golden,Kharkar,Yang}, \ourtool was initially evaluated using SpotBugs. To mitigate this threat and enhance the generalizability of our \revise{approach}, \ourtool is intentionally designed to depend only on the warning information which is common across different ASATs, such as the warning message~\cite{Sonarqube,PMD,Errorprone}. Additionally, even though the tools do not have some specific warning attributes, such as RANK, they can still be adapted to our \revise{approach} by simply ignoring these attributes, as they are not essential for the core functionality of \ourtool as shown in the \Cref{rq2}. This strategic decision ensures that our \revise{approach} can be broadly applicable and effective across a range of ASATs, facilitating its direct adaptation to other tools. Moving forward, our future work will expand to include a wider variety of ASATs, further broadening the scope and applicability of our approach. 

Another potential threat to internal validity is the generalizability of the evaluation in \Cref{rq4}. In this evaluation, we manually checked the warnings generated by \ourtool in the latest version of four popular open-source projects. However, our findings are limited to these specific projects. To address this, we selected widely used and highly starred projects on GitHub to cover a broad range of real-world scenarios. In addition, they are not part of the training set, which helps ensure that our model is not biased towards these projects. However, it is important to note that the results may not generalize to all projects.  

\revise{Furthermore, a threat to internal validity relates to the manual inspection process and the use of stratified random sampling. Since determining whether a warning corresponds to a real bug requires human judgment, the manual inspection is inherently subjective and susceptible to human error or bias.  To mitigate these threats, the manual inspection was conducted carefully by three senior developers, and any disagreements were resolved through discussion to reach a consensus. Additionally, the sample size was calculated to achieve a high confidence level, ensuring the statistical reliability of our findings.}

It is worth noting that we identified 25 real bugs among the 551 warnings retained by DeepFWI. In practice, a single bug may trigger multiple warnings, and the vast majority of static analysis warnings do not correspond to real bugs, making the identification of true bugs highly challenging. Our approach aims to maximize bug discovery by retaining all warnings related to each potential bug, rather than restricting to just one warning per bug. As a result, while the absolute precision is not very high, the concentration of true bugs among the filtered warnings is significantly higher than in the original set of 7,235 warnings produced by the static analysis tool. Additionally, there may be some bugs that were not detected during manual inspection. To further assess the reliability of our filtering, we conducted a sample analysis of the warnings filtered out as bug-insensitive. This sampling confirmed that the vast majority of these filtered warnings were indeed not related to real bugs, supporting the effectiveness of our filtering process.

\subsection{Limitations}
Our current instantiation of \ourtool targets Java projects analyzed by SpotBugs. Applying the approach to other programming languages (e.g., C/C++, C\#) or different ASATs would require adapting the parsing front-end, intermediate representation (e.g., LLVM IR or CIL instead of Jimple), and warning attributes, and then retraining on language-specific datasets. \revise{Moreover, the adaptation to another language also requires the generation of a dataset like BSWarnings for the respective language.} In addition, coding conventions and project-specific styles (e.g., use of fluent APIs, frameworks) may influence how informative warning messages and slices are. While the core idea of jointly modeling code, warning messages, and warning attributes is general, we cannot claim that the trained model will transfer directly to other languages or tools without such adaptation. We explicitly leave cross-language generalization and handling diverse coding standards as important directions for future work.

%% file: sections/6-relatedwork.tex
\section{related work}

\subsection{Learning-Based \revise{Approaches} for Warning Identification}
Most existing work on reducing false alarms focuses on using features involved in many aspects of warnings, code, and history information to predict warnings. Muske \etal provided a systematic survey of the post-processing of the warnings of ASATs~\cite{Muske}. For example,  Heckman \etal employed 51 features and evaluated 15 machine learning \revise{approaches} to achieve high accuracy ~\cite{Heckman}. Kim \etal used software change history to prioritize warning categories and proposed an algorithm for warning category prioritizing~\cite{rank2,rank3}. Liang \etal proposed 3 impact factors as features on a new heuristic for identifying actionable warnings~\cite{Liang}. Ruthruff \etal applied a screening \revise{approach} on existing logistic regression models to achieve a cost-effectively built prediction model to predict actionable warnings. Hanam et. al proposed a \revise{approach} to distinguish actionable warnings by finding alerts with similar code patterns~\cite{Hanam}. Based on these \revise{approaches} and features, Wang \etal proposed a  ``Golden Feature'' of 23 key attributes to achieve an AUC over 0.95 using linear SVM~\cite{golden,Yang}. However, Kang \etal found that there are leaked features and data duplication in the data, which can lead to a sudden drop in effectiveness~\cite{kang}. Moreover, Tu \etal tried to unlock the semi-supervised \revise{approach} for actionable warning identification. They proposed FRUGAL, a tuned semi-supervised \revise{approach} that executes on a simple optimization policy and outperforms the supervised learning \revise{approaches} for actionable warning identification. In addition, there is also another study to verify warnings at a function level. Kharkar \etal proposed a transformer-based learning approach to identify the false alarm on ASATs Infer~\cite{Kharkar}.
Our work differs from the above studies by focusing on a finer granularity and considering multi-modal semantics with deep learning.

Recently, Wen et al.~\cite{Wen} conducted a cutting-edge study using LLMs to investigate whether LLMs can serve as human experts in identifying ASAT warnings. However, their focus was limited to six types of bugs. Additionally, due to the high volume of warnings and thus the considerable time and financial expense associated with using models like ChatGPT/LLama2-70B (e.g., 30s/warning for LLama2), these \revise{approaches} may not be widely applicable to the large number of warnings typically encountered.

\subsection{Reanalysis-based \revise{Approaches} for Improving ASATs}
There are also studies working toward a different direction by improving the underlying analysis of ASATs. Many studies have attempted to reduce false positives in static analysis tools by focusing on pruning unreachable paths of a program. For example, Junker et al. proposed a \revise{approach} that treats static analysis as a model-checking problem and uses SMT-solvers to prune infeasible paths, reducing false positive warnings~\cite{Junker}. Nguyen et al. applied deductive verification to eliminate false alarms~\cite{nguyen2019multiple,nguyen2019reducing}. Li et al. introduced LLift, a framework that combines static analysis and LLMs to identify use-before-initialization (UBI) bugs within the Linux kernel~\cite{Li}. Ziyang Li et al. proposed a \revise{approach} that combines static analysis with symbolic execution to enhance the contextual analysis alleviating needs for human specifications and inspection~\cite{li2025iris}. 

Additionally, some \revise{approaches} employ dynamic analysis to enhance ASATs. For instance, Li et al. proposed residual investigation, a dynamic analysis technique to predict errors in executions and assess the validity of warnings~\cite{li2014residual}. Csallner et al. combined ASATs with concrete test-case generation to identify real warnings~\cite{csallner2005check}. Busse et al. combined ASATs with dynamic symbolic execution but found that the traces provided by ASATs do not effectively guide search~\cite{Busse}. Kallingal et al. proposed a syntactic patching algorithm to provide executable code for further testing~\cite{Kallingal}. Murali et al. introduced FuzzSlice, a framework that automatically prunes possible false positives among ASATs' warnings~\cite{Murali}.
Furthermore, some studies focus on simplifying manual inspection. These \revise{approaches} aim to screen warnings through semi-automated inspection~\cite{rival2005abstract,rival2005understanding}, feedback-based pruning~\cite{sadowski2015tricorder}, and checklist-based inspection~\cite{Panichella}.

%% file: sections/7-conclusion.tex
\section{Conclusion}
In this paper, we formulate the problem of warning identification at a fine-grained level, focusing on bug-sensitive warnings. To this end, we constructed \ourdataset, a dataset consisting of 280,273 static analysis warnings. Of these, 20,100 warnings are labeled as bug-sensitive, with the labeling process based on a combination of automated classification and validation through random sampling. To the best of our knowledge, this is the largest collection of data on warnings, ten times larger than the existing one~\cite{golden}. Then, we proposed \ourtool, a novel LSTM-based model for fine-grained warning identification of bugs for ASATs. \ourtool employs a multi-faceted representation which captures fine-grained semantics of source code and warnings from ASATs and focuses more on their correlations with cross-attention. With an extensive experiment, \ourtool shows an F1-score of 97.79\% for reducing false alarms and 67.06\% for confirming effective warnings. Compared with baselines, \ourtool outperforms 9.03\% in precision, and 13.17\% in F1-score. Moreover, we employed \ourtool on four real-world projects to filter out about 92\% of warnings and then found 25 new bugs with minimal manual effort. We have released our data on our website~\cite{website} to replicate the results of this work and encourage further research.